\documentclass[preprintnumbers,amsmath,amssymb,floatfix,11pt,prd,onecolumn,
superscriptaddress,nofootinbib]{revtex4}
\usepackage{latexsym}
\usepackage{epsfig}
\usepackage{amsmath}
\usepackage{epstopdf}
\usepackage{amssymb}
\usepackage{graphicx}
\usepackage{hyperref}
\usepackage{tabularx}
\usepackage{varioref}
\usepackage{placeins}
\newcolumntype{C}[1]{>{\centering\arraybackslash}p{#1}}

\begin{document}
\title{\bf Black Holes and White Holes as Particle Accelerators}

\author{Ayesha Zakria}
\email{ayesha.zakria@bcoew.edu.pk}
\affiliation{Department of Mathematics, Bilquis Post Graduate College For Women, Air University, Islamabad}

\author{Qurat-ul-Ain Satti}
\email{qubee646@gmail.com}
\affiliation{Department of Mathematics, Bilquis Post Graduate College For Women, Air University, Islamabad}

\maketitle
\begin{center}
\textbf{\textbf{ABSTRACT}}
\end{center}

 We investigate particle collisions in non-extremal black hole that can probably induced to extremely high center of mass energy $E_{\text{cm}}$. We consider the collision of two particles where first particle comes from far to the outer horizon of the Reissner-Nordstr\"{o}m black hole and second particle emanates from the white hole region. It is exhibited that unbounded $E_{\text{cm}}$ requires that second particle lapse near the bifurcation point. We discuss the collision of particles close to the outer horizon in detail.

 \newpage

\section{Introduction}
Many years ago, it was noticed that if two particles collide close to a rotating extremal black hole, the energy in the center of mass frame $E_{\text{cm}}$ can be unbounded \cite{1}. This effect was named as Ba\~{n}ados-Silk-West (BSW) effect. Afterward, this effect was generalized to three cases: non-extremal black holes \cite{2}, generic rotating black holes \cite{3} and nonrotating charged black holes \cite{4}. In all three cases, it is indicated that both particles move towards the horizon of the black holes.

Meantime, there are some scenarios for head-on-collision in which one of the two particles moves in another direction from the horizon. It's detail was mentioned in \cite{5} but the term "white hole" was not used. The comprehensible treatment of this type of scenario was described in \cite{6} where the aspect of white holes was emphasised and it was observed that unbounded $E_{\text{cm}}$ takes place for the Schwarzschild metric. Since the spacetime of the eternal black hole includes unavoidably two regions; black hole and white hole. According to the scenario mentioned in \cite{6}, first particle progresses towards the future horizon and second particle accesses the past horizon from the inner white hole region. In terms of R- and T-regions \cite{7}, first particle travels within $R-$ region and second particle crosses from the expanding $T-$region to the $R-$region. Unlike the typical BSW effect where fine tuning between parameters of one of the two particles is required, this scenario works for generic particles and eternal black holes.

The presence of white holes is controversial. Especially, thay can be instable \cite{8}. Several years ago, an interesting speculation was explained which conclude that white holes can act as region retarded in the expansion of surrounded matter in Universe \cite{9}. The energetics of white holes has been elaborated in a different circumstances \cite{11}. The framework of spacetime includes interchange of $R-$ and $T-$ regions, for example, this occurs for black universes \cite{10} and the motion of self-gravitating shells \cite{12}. The energy $E_{\text{cm}}$ tends to be different for two colliding particles  at the inner horizon of a non-extremal Kerr black hole \cite{13,14}. The high scattering energy of particles can be acquired for an extremal and non-extremal Kerr black hole \cite{15}.

There is a possibility of obtaining unbounded energy $E_{\text{cm}}$ when particles collide close to the inner horizon of the black hole \cite{16} when both particles move (i) in the same direction or (ii) in the opposite direction. The energy $E_{\text{cm}}$ can grow unbounded in the local area of any bifurcation surface \cite{17}. This effect can be understood with the help of a simple comparison with particle collisions in flat space-time \cite{18}. The energy $E_{\text{cm}}$ for two particles in the background of a Kerr-Newman-Taub-NUT \cite{19} and Kerr-MOG black holes \cite{20} has been investigated.

The theory of hight energy physics is incomplete without discussion of collision of particles close to the white holes. The master plan of particle collisions in the metric of a non-extremal black hole that can potentially lead to extremely high energy $E_{\text{cm}}$ \cite{21} where first particle appears from infinity to the black hole horizon and second particle arrives from a white hole region. We will further extend this work when first particle comes from infinity to the outer horizon of a Reissner-Nordstr\"{o}m black hole and second particle emanates from a white hole region.

Reissner-Nordstr\"{o}m (RN) metric describes the geometry of the spacetime that surrounds a non-rotating charged spherical black hole. In reality, a highly charged black hole would be quickly neutralized by interactions with matter in its vicinity and therefore such solution is not extremely relevant to realistic astrophysical situations. Nevertheless, charged black holes illustrate a number of important features of more general situations. The RN metric is reduced to Schwarzschild metric in the absence of charge.

In the present work, we will examine the main features for RN metric. We will also show the basic equations for a pure radial motion in equitorial plane.  We will use some important factors to describe the BSW effect and its modification in our paper. We use the system of units in which constants $G=c=1$ throughout this paper.
\section{EQUATIONS OF MOTION}
\begin{figure}[h!]
\centering
\includegraphics[width=8cm, height=8cm]{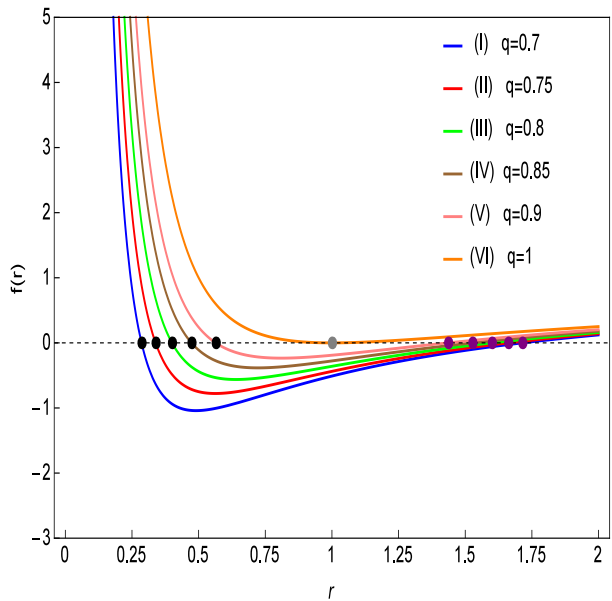}
\caption{The function $f(r)$ for different values of charge $q$. Black points represent outer horizons and purple points show inner horizons of the non-extremal RN black hole. Gray point identifies horizon of the extremal RN black hole.}\label{Cf1}
\end{figure}

Let us consider the metric
\begin{equation}\label{1}
ds^{2}=-f(r)dt^{2}+\frac{dr^{2}}{f(r)}+r^{2}{d\theta^{2}+\sin^{2}\theta d\phi^{2}}.
\end{equation}
It is the metric of the eternal hole where $f(r_{\pm})=0$. For RN metric $f(r)=\Big(1-\frac{r_{+}}{r}\Big)\Big(1-\frac{r{-}}{r}\Big)$ and $r_{\pm}=\mu\pm\sqrt{\mu^{2}-q^{2}}$. The RN metric depends upon the charge $q$ and mass $\mu$. Now we consider the radial motion. By using Euler langrange formalism and normalization condition, one can find the equations of motion in the equatorial plane
\begin{eqnarray}
u^{r}&=&\epsilon\sqrt{E^{2}-f},\label{2}\\
u^{t}&=&\frac{E}{f},\label{3}
\end{eqnarray}                                                                                                                                                where $u^{\nu}=$ {\Large $\frac{dx^{\nu}}{d\tau}$}, $\tau$ is the proper time and for pure radial motion $u^{\phi}=0$. So from (\ref{2}) and (\ref{3}), we have
\begin{equation}\label{4}
\frac{dr}{dt}=\epsilon\frac{f\sqrt{E^{2}-f}}{E},
\end{equation}
where $E=${\Large $\frac{\mathcal{E}}{m}$}, $\mathcal{E}$ is the energy, $E$ is the specific energy, $m$ is the mass of free particle and $\epsilon=\pm1$ depending on the direction of motion. Using Eq. (\ref{2}), one finds
\begin{equation}\label{A4}
(u^{r})^{2}+f(r)=E^{2}.
\end{equation}
Clearly, it is form of an `energy' equation, in which the function $f(r)$ plays the role of an effective potential. The properties of the radial trajectories can be obtained directly from Eq. (\ref{A4}) by plotting the function $f(r)$ for different values of charge $q$. The plots are shown in Figure $\ref{Cf1}$.
Black points represent place of residence of the outer horizons and purple points show place of residence of the inner horizons of the non-extremal RN black hole. Gray point identifies place of residence of the horizon of the extremal RN black hole.
\section{CENTER OF MASS ENERGY}
 Now let us consider the two colliding particles, then energy in the center of mass frame is given by
\begin{equation}\label{5}
E^{2}_{\text{cm}}=-P_{\mu}P^{\mu},
\end{equation}
$P_{\mu}=m_{1}u_{1\mu}+m_{2}u_{2\mu}$, where $P_{\mu}$, $u_{\mu}$ and $m$ are the 4-momentum, 4-velocity and rest mass of the colliding particles, respectively. Here $u_{i\mu}u_{i}^{\mu}=-1$ and $ -u_{1\mu}u_{2}^{\mu}=\gamma$, then the Eq. (\ref{5}) becomes
\begin{equation}\label{6}
E^{2}_{\text{cm}}=m_{1}^{2}+m_{2}^{2}+2m_{1}m_{2}\gamma,
\end{equation}
where
\begin{equation}\label{A6}
\gamma=\frac{E_{1}E_{2}-\epsilon_{1}\epsilon_{2}\sqrt{E_{1}^{2}-f}\sqrt{E_{2}^{2}-f}}{f},
\end{equation}
is the Lorentz factor of relative motion.
On considering the collision of two particles, let first particle with $\epsilon_{1}=-1$ and second particle with $\epsilon_{2}=+1$ collide at $r=r_{c}$. Then, the equations of motion give
\begin{equation}\label{7}
E^{2}_{\text{cm}}\big|_{r\rightarrow r_{c}}=m_{1}^{2}+m_{2}^{2}+2m_{1}m_{2}\gamma|_{r\rightarrow r_{c}},
\end{equation}
where
\begin{equation}\label{A7}
\gamma|_{r\rightarrow r_{c}}=\frac{E_{1}E_{2}+\sqrt{E_{1}^{2}-\Big(1-\frac{r_{+}}{r_{c}}\Big)\Big(1-\frac{r{-}}{r_{c}}\Big)}\sqrt{E_{2}^{2}-\Big(1-\frac{r_{+}}{r_{c}}\Big)\Big(1-
\frac{r{-}}{r_{c}}\Big)}}{\Big(1-\frac{r_{+}}{r_{c}}\Big)\Big(1-\frac{r{-}}{r_{c}}\Big)}.
\end{equation}
If the collision occurs close to the outer horizon, then $r_{c}\rightarrow r_{+}$, and also
\begin{equation}\label{B7}
f_{c}=f(r_{c})=\Big(1-\frac{r_{+}}{r_{c}}\Big)\Big(1-\frac{r{-}}{r_{c}}\Big)\rightarrow 0.
\end{equation}
Therefore, we attain Eq. (\ref{7}) diverges. Unlike BSW effect, this effect consists of the future (black hole) horizons and the past (white hole) horizons.
\section{KRUSKAL COORDINATES}
Now to describe ingoing and outgoing null geodesics, consider a metric mentioned in Eq. (\ref{1}). For a radially moving photons $ds=d\theta=d\phi=0$, one can find the following expression                                                                                                                                                   \begin{equation}\label{8}                                                                                                                                   dt=\pm\frac{r^{2}}{(r-r_{-})(r-r_{+})}dr,                                                                                                                     \end{equation}                                                                                                                                              integration yields $t=\pm r^{*}+c$, where $r^{*}$ is the tortoise co-ordinate given by                                                                                                      \begin{equation}\label{10}                                                                                                                                r^{*}=\int^{r}\frac{dr}{f}.                                                                                                                                      \end{equation}
To make more ascent, let us introduce a Kruskal co-ordinate that describes the whole spacetime which includes both the black and the white
 hole regions. We use the co-ordinates $U=-\exp(-\kappa u)$, $V=\exp(\kappa v)$ in the $R-$region $r>r_{+}$ where $u=t-r^{\ast}$, $v=t+r^{\ast}$ and $\kappa$ is the surface gravity. Using $U$, $V$ and Eq. (\ref{10}), one can find a relation
\begin{eqnarray}
UV&=&-\exp({2\kappa r^{\ast}}),\label{13}\\
\frac{V}{|U|}&=&\exp(2\kappa t). \label{A16}
\end{eqnarray}
Close to the outer horizon,
\begin{equation}\label{B16}
r^{\ast}\approx\frac{1}{2r_{+}^{2}\kappa_{+}}\bigg(r_{+}^{2}\ln\bigg|\frac{r}{r_{+}}-1\bigg|-r_{-}^{2}\ln\Big|\frac{r}{r_{-}}-1\Big|\bigg)+A,
\end{equation}
where $A$ is a constant of integration, and
\begin{equation}\label{B16}
f(r)\approx2\kappa_{+}r_{+}UV\bigg(\frac{r}{r_{-}}-1\bigg)^{\frac{r_{-}^{2}}{r_{+}^{2}}}.
\end{equation}
For the RN metric, using the constant of integration properly, we get the exact results
\begin{eqnarray}
r^{\ast}&=&r-\frac{r_{-}^{2}}{r_{+}-r_{-}}\ln\bigg|\frac{r}{r_{-}}-1\bigg|+\frac{r_{+}^{2}}{r_{+}-r_{-}}\ln\bigg|\frac{r}{r_{+}}-1\bigg|,\label{17}\\
\kappa_{\pm}&=&\frac{r_{\pm}-r_{\mp}}{2r_{\pm}^{2}},\label{18}\\
f&=&\frac{r_{+}r_{-}}{r^{2}}\exp\bigg(-\frac{r}{r_{+}r_{-}}\bigg)UV.\label{19}
\end{eqnarray}                                                                                                                                                In terms of $U$ and $V$ co-ordinates, the metric (\ref{1}) takes the form
\begin{equation}\label{20}
ds^{2}=-FdUdV+r^{2}(d\theta^{2}+\sin^{2}\theta d\phi^{2}),
\end{equation}
where
\begin{equation}\label{21}
F=f\frac{du}{dU}\frac{dv}{dV}=\frac{f}{k^{2}UV}.
\end{equation}
Now, let us consider a collision point $U_{1}=U_{2}, V_{1}=V_{2}$ and in terms of $t$ co-ordinates, we have $t_{1}(r)=t_{2}(r)$. Now using Eq. (\ref{4}), we have
\begin{equation}\label{25}
t_{1}(r)=E_{1}\int^{r_{1}}_{r}\frac{dr}{f\sqrt{E_{1}^{2}-f}}.
\end{equation}                                                                                                                                                 During the motion of first particle, $r_{1}$ is the starting point. So as a result $t_{1}(r_{1})=0$. Similarly, for second particle
\begin{equation}\label{26}
t_{2}(r)=t_{1}(r_{c})-E_{2}\int^{r_{c}}_{r}\frac{dr}{f\sqrt{E_{2}^{2}-f}}.
\end{equation}
Clearly, $t_{1}(r_{c})=t_{2}(r_{c})$. If first particle comes from infinity and $E\geq1$, then equations of motion in terms of $(u,v)$ coordinate system follow from Eqs. (\ref{2}), (\ref{3}) and (\ref{4}). They become
\begin{eqnarray}
\frac{du}{dr}&=&\frac{\epsilon}{\sqrt{E^{2}-f}\big(E+\epsilon\sqrt{E^{2}-f}\big)},\label{A28}\\
\frac{dv}{dr}&=&\frac{\epsilon E+\sqrt{E^{2}-f}}{f\sqrt{E^{2}-f}}.\label{A29}
\end{eqnarray}
Further, the equations of motion in terms of Kruskal coordinates read
\begin{eqnarray}
\frac{dU}{dr}&=&-\frac{\epsilon \kappa U}{\sqrt{E^{2}-f}\big(E+\epsilon\sqrt{E^{2}-f}\big)},\label{28}\\
\frac{dV}{dr}&=&\frac{\kappa V}{\sqrt{E^{2}-f}\big(\epsilon E-\sqrt{E^{2}-f}\big)}.\label{29}
\end{eqnarray}
\section{kinematics}
Before the collision of two particles, first particle will cross the future horizon when $U=0$ and $V=V_{1}$, and second particle will cross the past horizon when $U=U_{2}$ and $V=0$. These particles collide at the intermediate point with $|U_{c}|=O(1)$ and $V_{c}=O(1)$, both equations offer a finite $\gamma$. We must set up the collision very close to the outer horizon to get large center of mass energy, where $f_{c}$ is extremely small and $\gamma$ is infinite followed by Eq. (\ref{A7}). As we are interested in the effects near the white hole horizon $V=0$, we require $V_{c}\ll1$. This involves effects for the residences of a trajectory of both particles.
\subsection{First Particle}\label{Subsec:B1}
\hspace{1cm}Let us assume first particle started its movement at $t_{1}=0$. We consider, for $t<0$, it remained at rest and $r=r_{1}=$ constant, then $t$ is greater than zero on its further trajectory. It is clear that $|U_{c}|<V_{c}$ by Eq. (\ref{A16}). The collision occurs close to the bifurcation point $U=V=0$, as we have both $|U_{c}|\ll1$ and $V_{c}\ll1$. Due to the disagreement of $V_{c}\ll1$ and Eq. (\ref{A16}), we can say that close to the generic point of the white hole horizon where $U=O(1)$ and $V=0$, collision can not take place.
\subsection{Second Particle}\label{Subsec:B2}
\hspace{1cm}By assumption, second particle moves from a white hole with $\epsilon_{2}=+1$. We wish collision occurs close to the outer horizon of the white hole. The term $r_{c}-r_{+}$ is small. Thus, we derive
\begin{equation}\label{30}
U_{c}\approx U_{+}+\bigg(\frac{dU}{dr}\bigg)_{+}(r_{c}-r_{+})\approx U_{+}-\frac{\kappa_{+}U_{+}}{2E_{2}^{2}}(r_{c}-r_{+})\approx U_{+}\bigg(1-\frac{f_{c}}{4E_{2}^{2}}\bigg),
\end{equation}
where $U_{+}=U(r_{+})$ and close to the outer horizon
\begin{equation}\label{31}
f(r)\approx2\kappa_{+}(r-r_{+}).
\end{equation}
Any finite specific energy of second particle $E_{2}$ offers a small correction to $U_{+}$. So $U_{c}\approx U_{+}$. Hence the point where second particle intercross the horizon and the point of collision are situated close to the bifurcation point. We are able to add the case where both $E_{2}^{2}$ and $f_{c}$ are small and have the same order, i.e.,
\begin{equation}\label{32}
E_{2}^{2}\sim f_{c},
\end{equation}                                                                                                                                               so, Eq. (\ref{30}) is not useful. For second particle close to the outer horizon, Eqs. (\ref{28}) and (\ref{31}) give
\begin{equation}\label{33}
\frac{d}{dr}\ln|U|\approx-\frac{\kappa_{+}}{\sqrt{E_{2}^{2}-2\kappa_{+}(r-r_{+})}}\frac{1}{\sqrt{E_{2}^{2}-2\kappa_{+}(r-r_{+})}+E_{2}}.
\end{equation}
It is easy to use $E_{2}^{2}=2\kappa_{+}(r_{0}-r_{c})$. However, $r_{0}$ is close to $r_{c}$ which is successively close to $r_{+}$. Collision should occur earlier then second particle reaches the turning point, in any other case, $\epsilon_{2}$ will change the sign and head-on collision will not arise. Therefore, $r_{c}\leq \frac{r_{+}+r_{0}}{2}$.
Now, after integration of Eq. (\ref{33}) with boundary conditions $U(r_{c})=U_{c}$, we can find the relation
\begin{equation}\label{37}
U\approx\frac{U_{c}(\sqrt{s}+\sqrt{s-x}}{\sqrt{s}+\sqrt{s-x_{c}}},
\end{equation}
where
\begin{eqnarray}
x&=&\Big(\frac{r}{r_{+}}-1\Big)\Big(1-\frac{r_{-}}{r_{+}}\Big),\label{38}\\
s&=&\Big(\frac{r_{0}-r_{c}}{r_{+}}\Big)\Big(1-\frac{r_{-}}{r_{+}}\Big),\label{39}
\end{eqnarray}
and non-negative radicals require $x_{c}\leq s$, $x_{c}=x(r_{c})$, thus
\begin{equation}\label{40}
U_{+}\approx \frac{2U_{c}}{1+\sqrt{1-\frac{x_{c}}{s}}}.
\end{equation}
We see that $U_{c}$ and $U_{+}$ are small with reason given above, also both have the same order. Therefore collision takes place close to the bifurcation point for the both cases $E_{2}\gg f_{c}$ and $E_{2}\sim f_{c}\ll1$. Note that if $E_{2}\sim\sqrt{f_{c}}$, $E_{1}=O(1)$, $\gamma=O\Big(f_{c}^{-\frac{1}{2}}\Big)$, so increase of $E_{\text{cm}}$ is slower than in the case $E_{2}\gg \sqrt{f_{c}}$ where $\gamma=O\big(f_{c}^{-1}\big)$. When both particles have energies $E_{1}\sim E_{2}\sim\sqrt{f_{c}}$ then the effect of high energy collision does not take place.

\begin{figure}[h!]
\centering
\includegraphics[width=8cm, height=8cm]{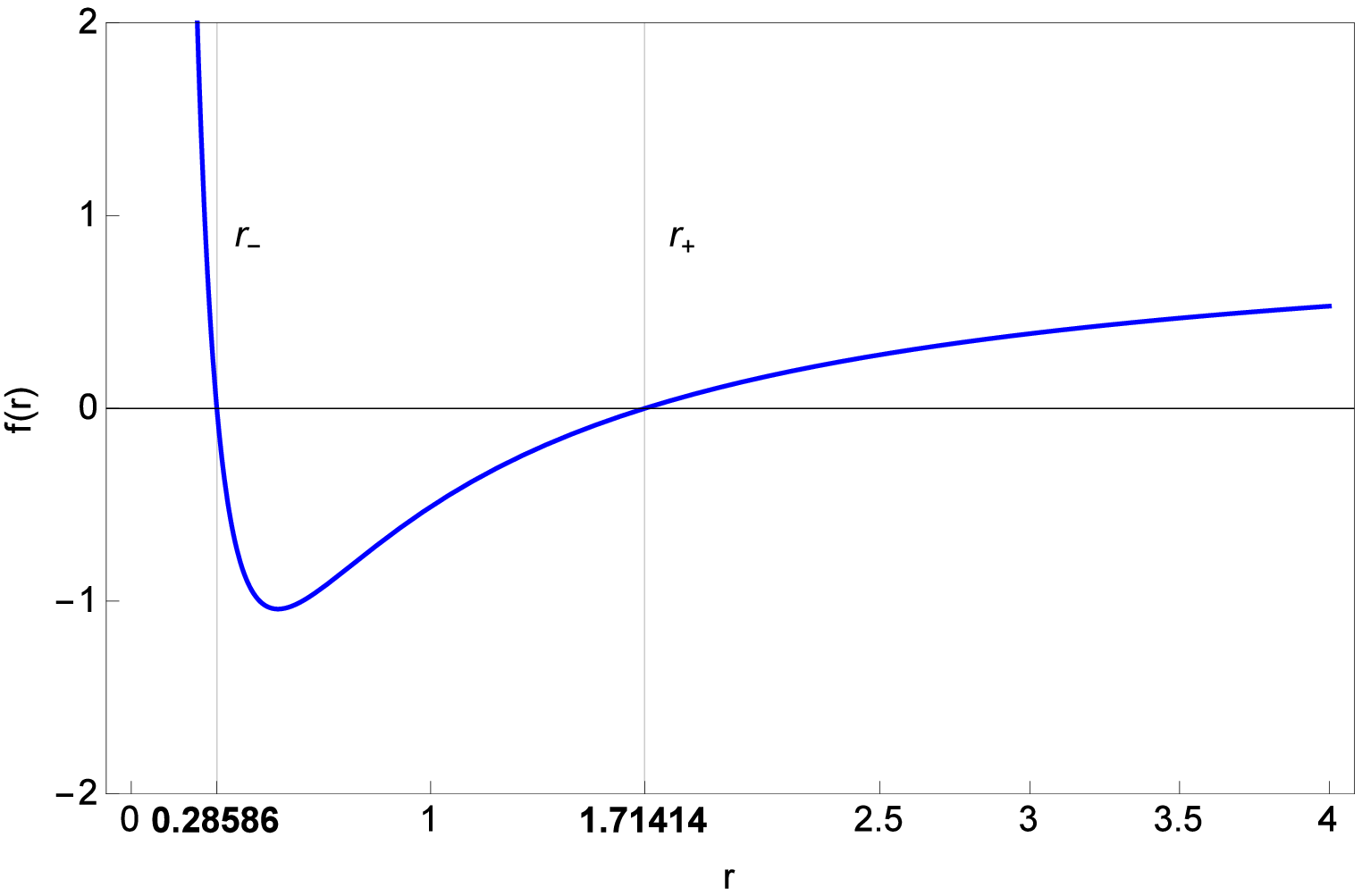}
\includegraphics[width=8cm, height=8cm]{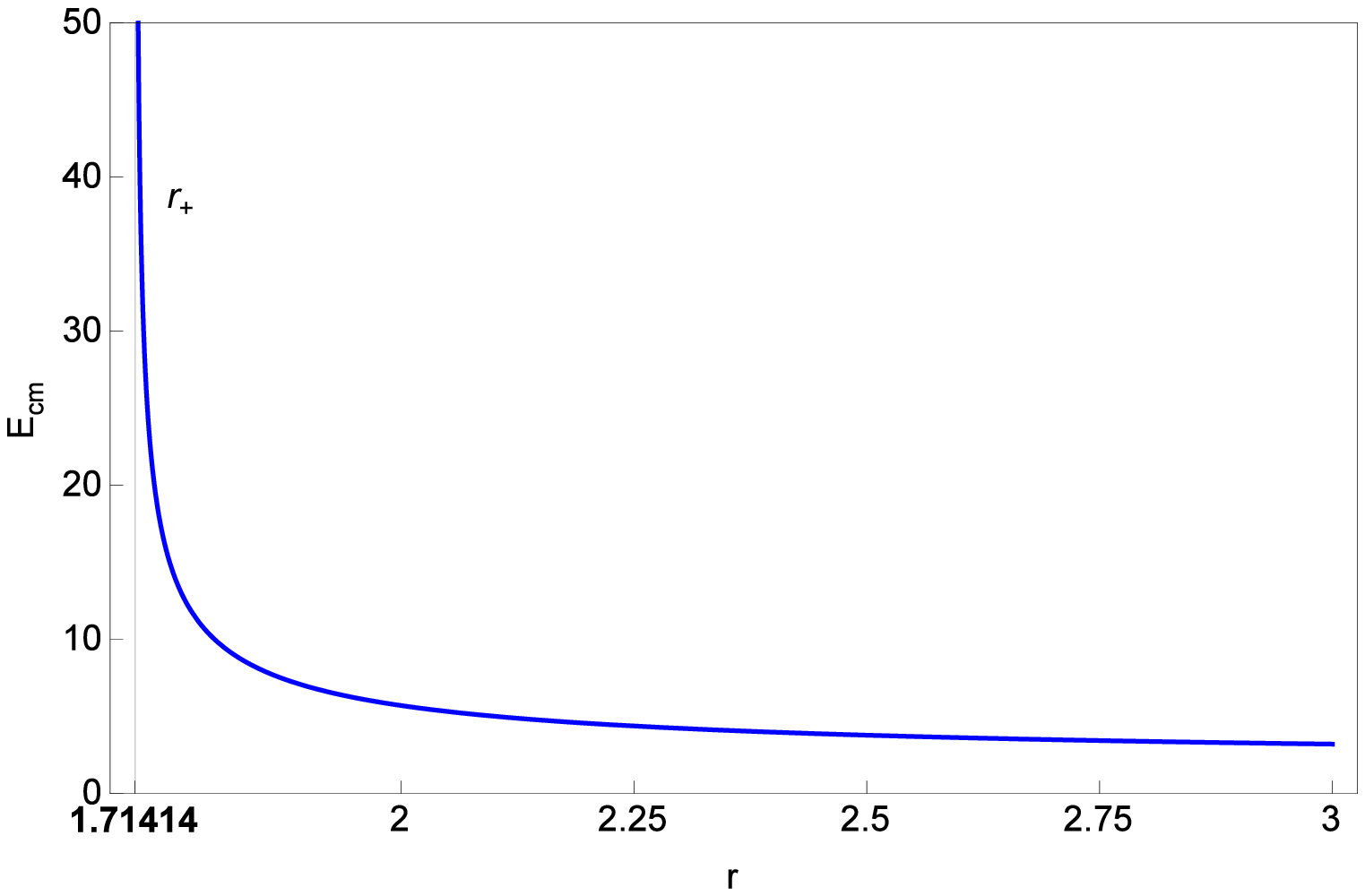}
\caption{The function $f(r)$ (left figure) and $E_{\text{cm}}$ (right figure) where first particle comes from far to the outer horizon of the non-extremal RN black hole with specific energy $E_{1}=1$ and second particle emanates from the white hole region with specific energy $E_{2}=1$. We set $\mu=1$, $m_{1}=m_{2}=1$ and $q=0.7$. Vertical lines recognize place of residence of the outer and inner horizons.}\label{CMEf1}
\end{figure}
\begin{figure}[h!]
\centering
\includegraphics[width=8cm, height=8cm]{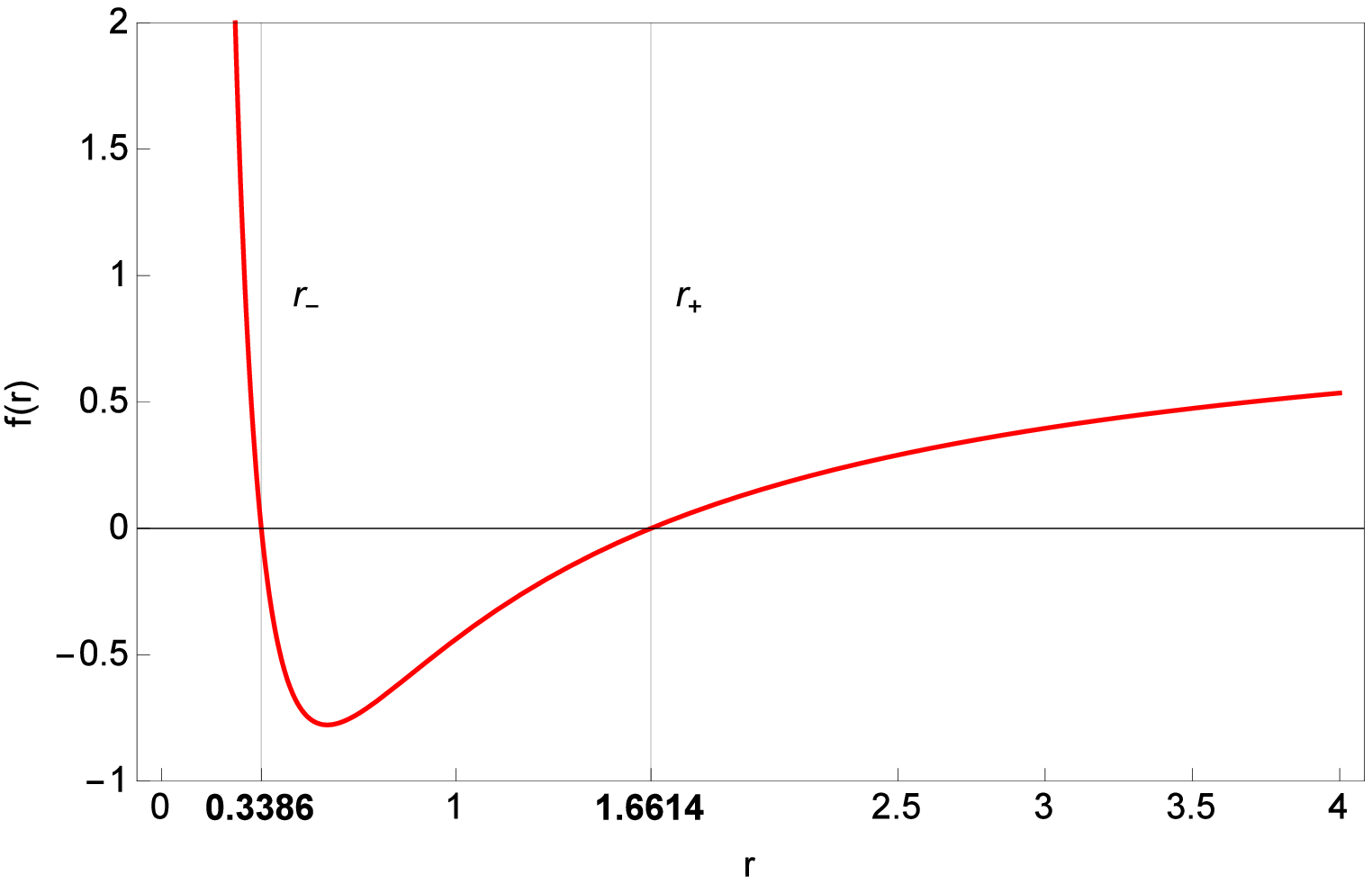}
\includegraphics[width=8cm, height=8cm]{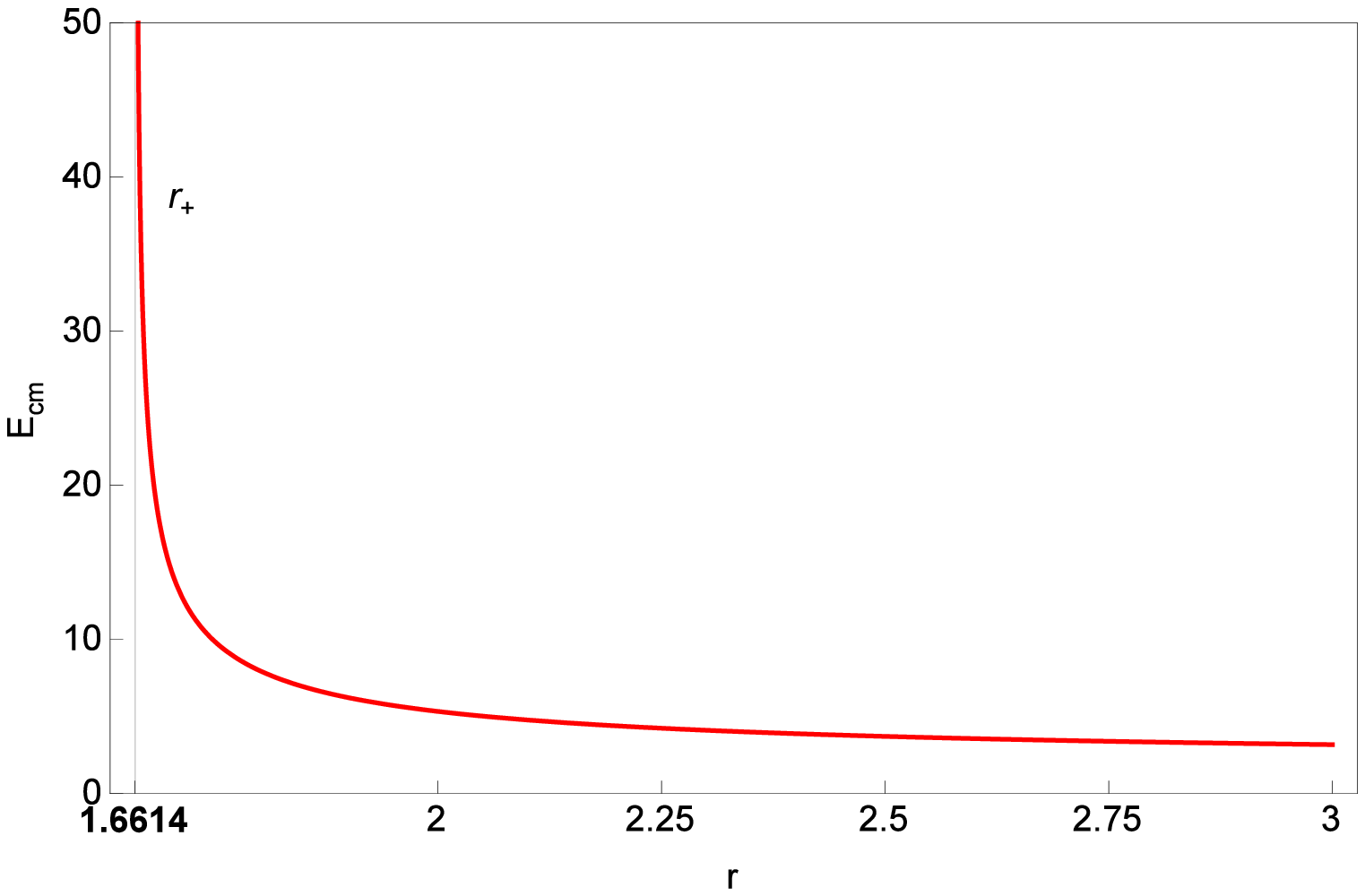}
\caption{The function $f(r)$ (left figure) and $E_{\text{cm}}$ (right figure) where first particle comes from far to the outer horizon of the non-extremal RN black hole with specific energy $E_{1}=1$ and second particle emanates from the white hole region with specific energy $E_{2}=1$. We set $\mu=1$, $m_{1}=m_{2}=1$ and $q=0.75$. Vertical lines recognize place of residence of the outer and inner horizons.}\label{CMEf2}
\end{figure}
\begin{figure}[h!]
\centering
\includegraphics[width=8cm, height=8cm]{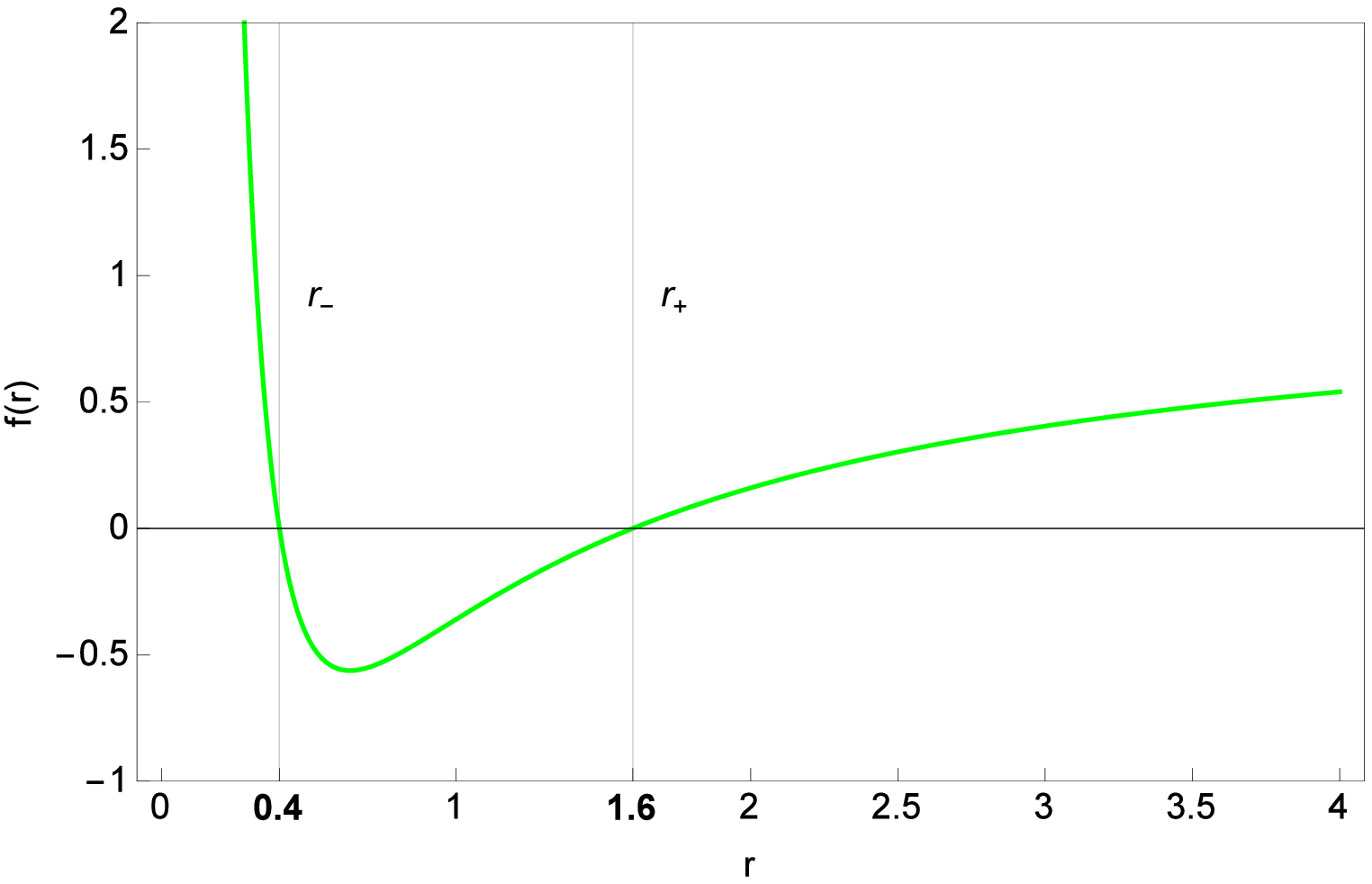}
\includegraphics[width=8cm, height=8cm]{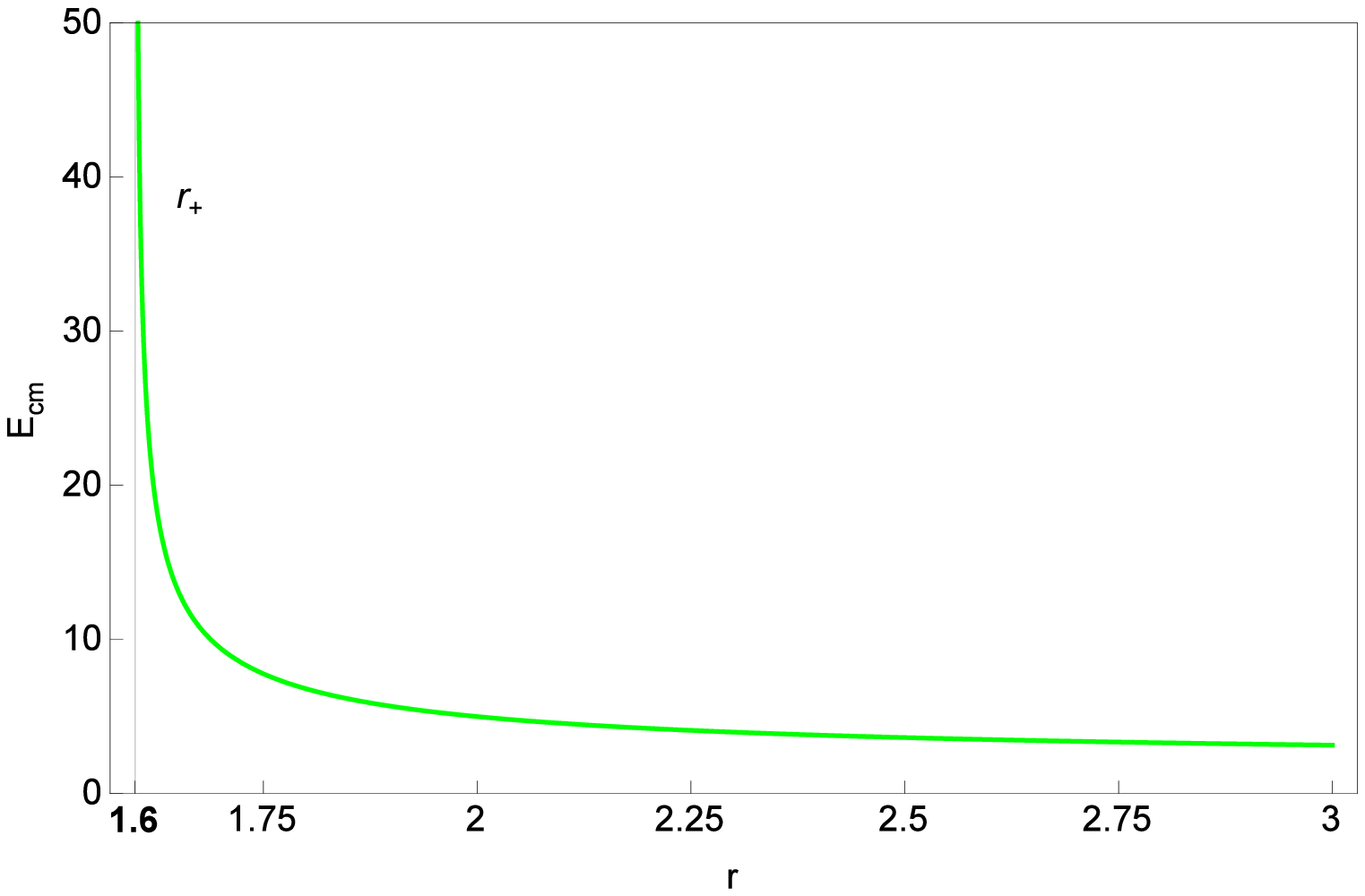}
\caption{The function $f(r)$ (left figure) and $E_{\text{cm}}$ (right figure) where first particle comes from far to the outer horizon of the non-extremal RN black hole with specific energy $E_{1}=1$ and second particle emanates from the white hole region  with specific energy $E_{2}=1$. We set $\mu=1$, $m_{1}=m_{2}=1$ and $q=0.8$. Vertical lines recognize place of residence of the outer and inner horizons.}\label{CMEf3}
\end{figure}
\begin{figure}[h!]
\centering
\includegraphics[width=8cm, height=8cm]{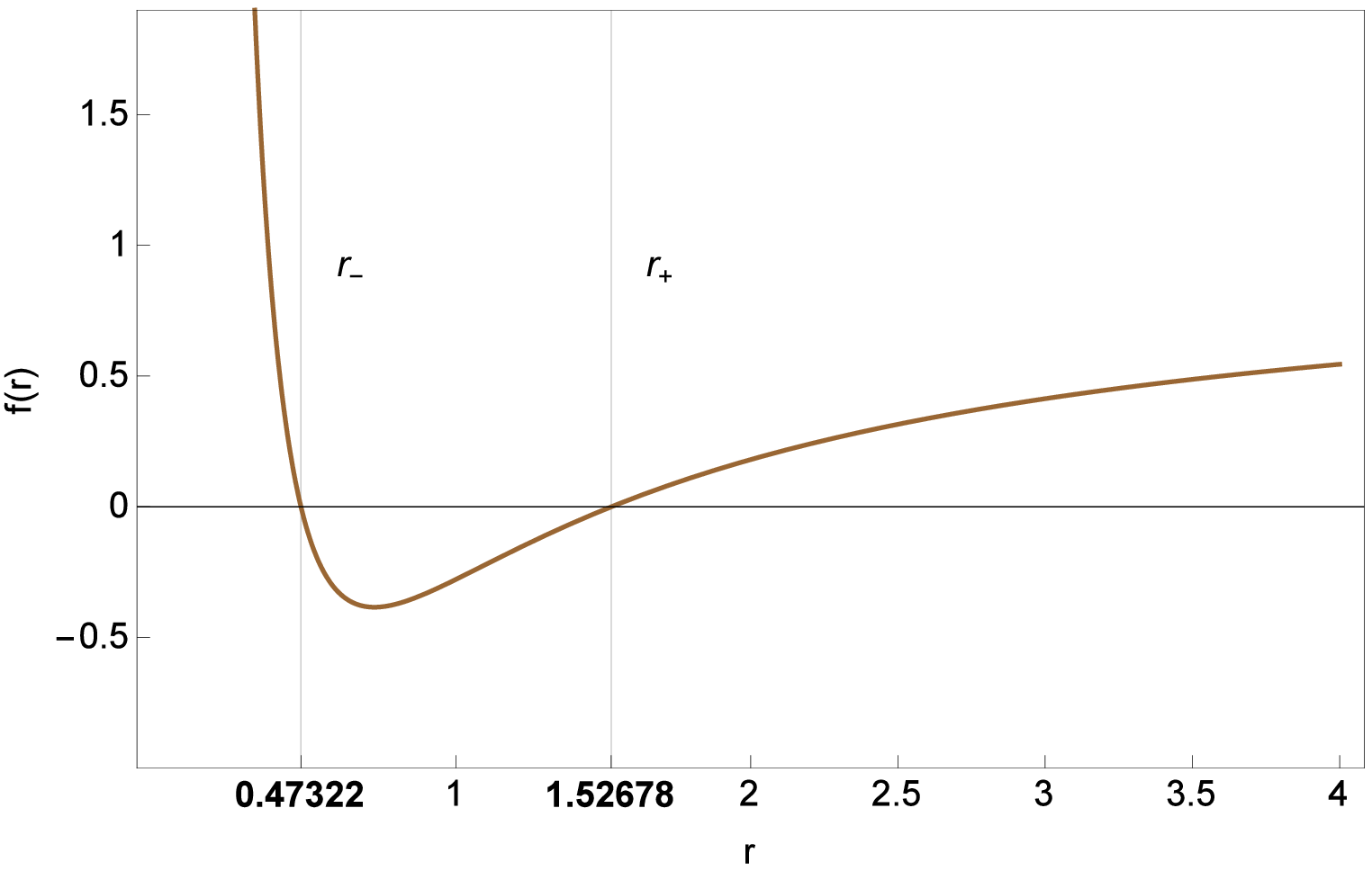}
\includegraphics[width=8cm, height=8cm]{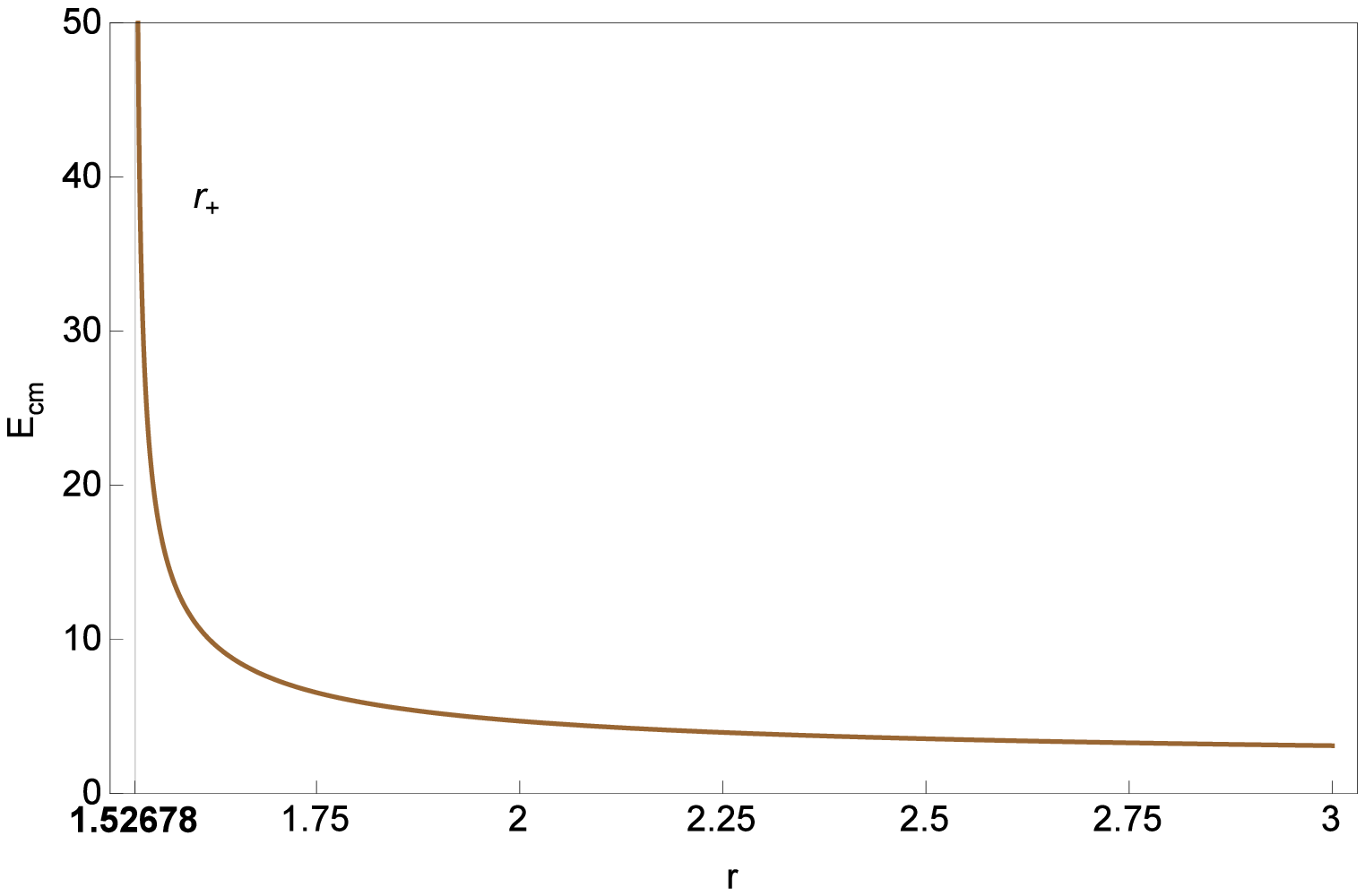}
\caption{The function $f(r)$ (left figure) and $E_{\text{cm}}$ (right figure) where first particle comes from far to the outer horizon of the non-extremal RN black hole with specific energy $E_{1}=1$ and second particle emanates from the white hole region  with specific energy $E_{2}=1$. We set $\mu=1$, $m_{1}=m_{2}=1$ and $q=0.85$. Vertical lines recognize place of residence of the outer and inner horizons.}\label{CMEf4}
\end{figure}
\begin{figure}[h!]
\centering
\includegraphics[width=8cm, height=8cm]{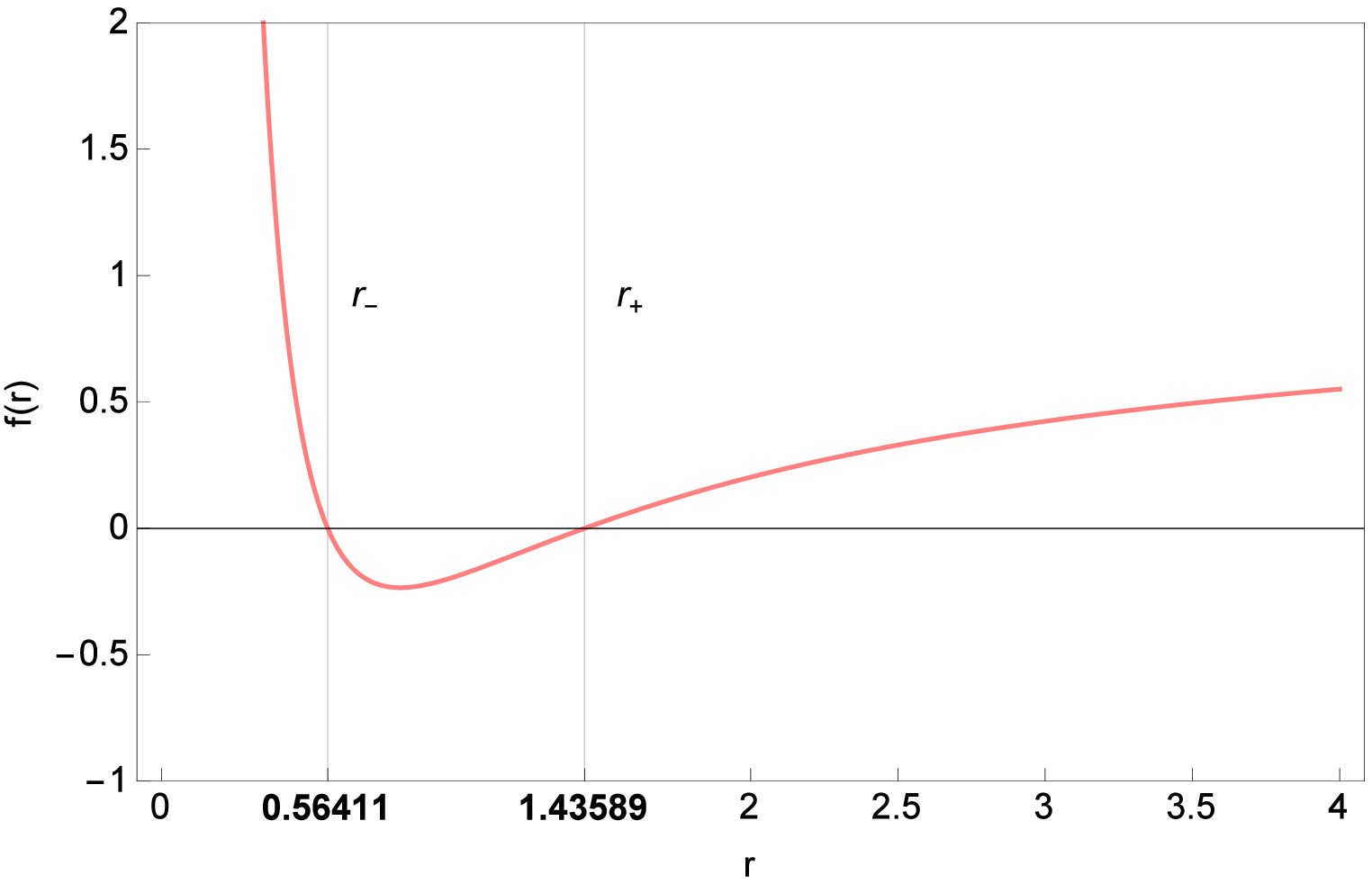}
\includegraphics[width=8cm, height=8cm]{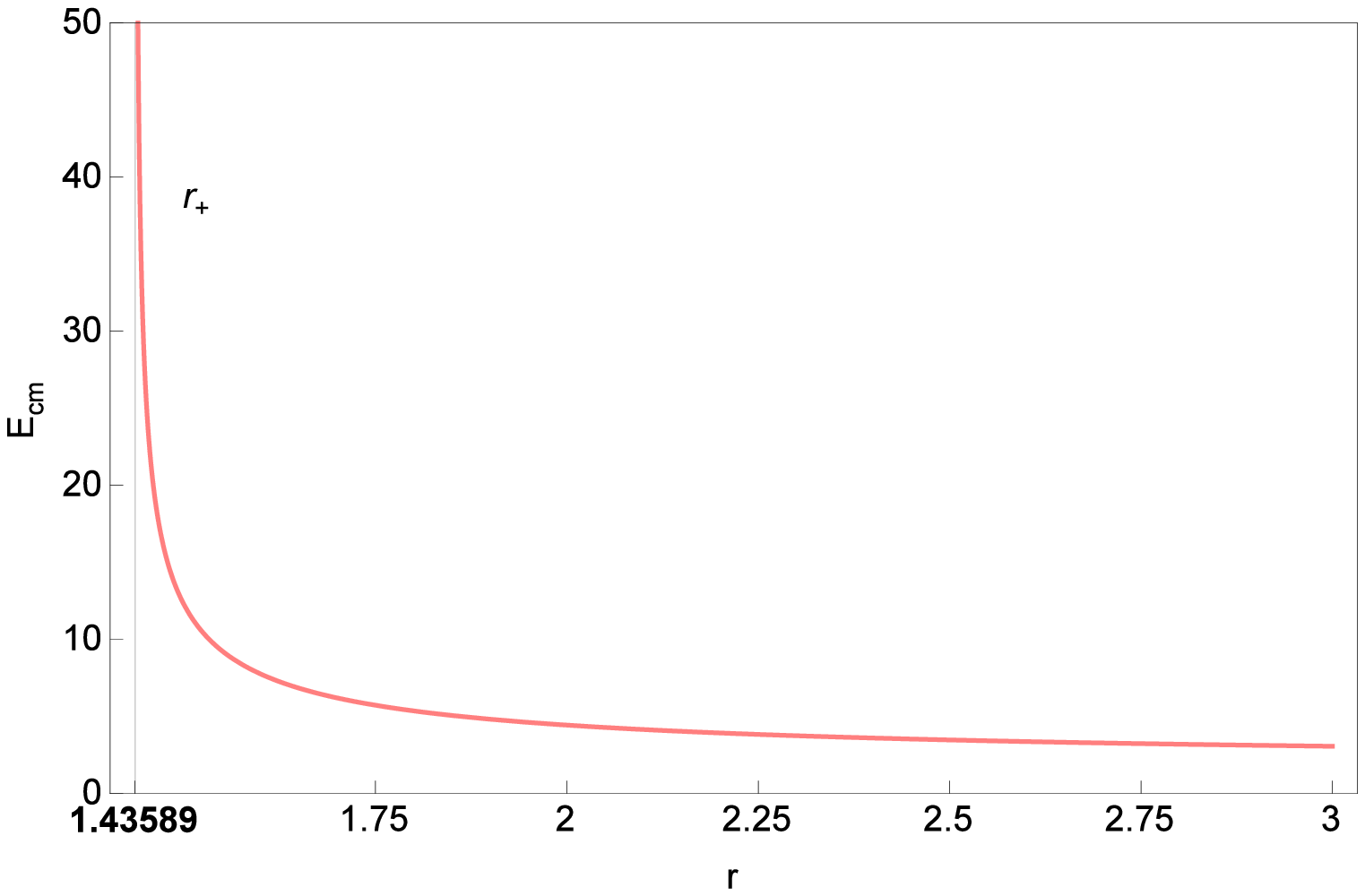}
\caption{The function $f(r)$ (left figure) and $E_{\text{cm}}$ (right figure) where first particle comes from far to the outer horizon of the non-extremal RN black hole with specific energy $E_{1}=1$ and second particle emanates from the white hole region  with specific energy $E_{2}=1$. We set $\mu=1$, $m_{1}=m_{2}=1$ and $q=0.9$. Vertical lines recognize place of residence of the outer and inner horizons.}\label{CMEf5}
\end{figure}
\begin{figure}[h!]
\centering
\includegraphics[width=8cm, height=8cm]{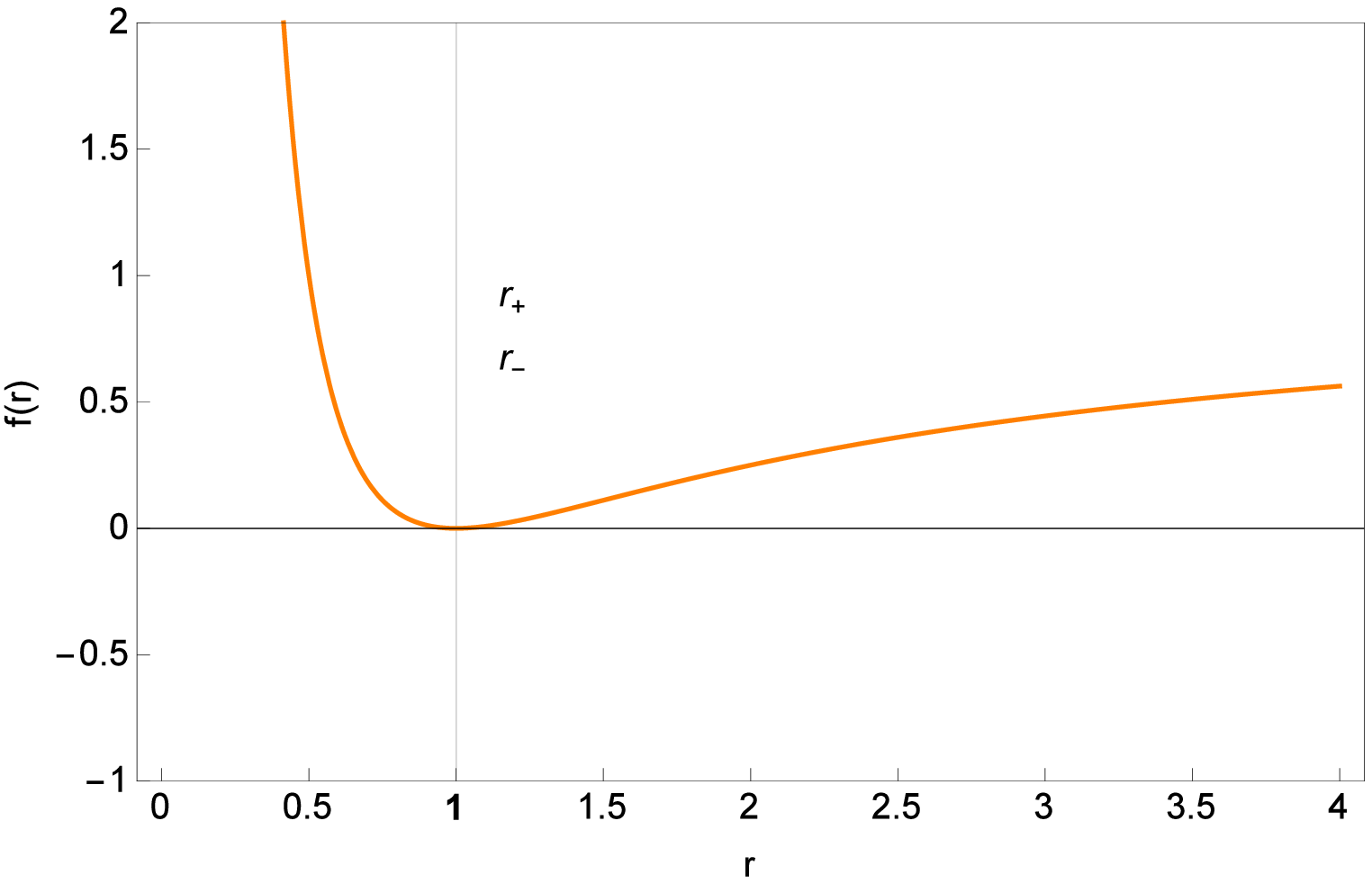}
\includegraphics[width=8cm, height=8cm]{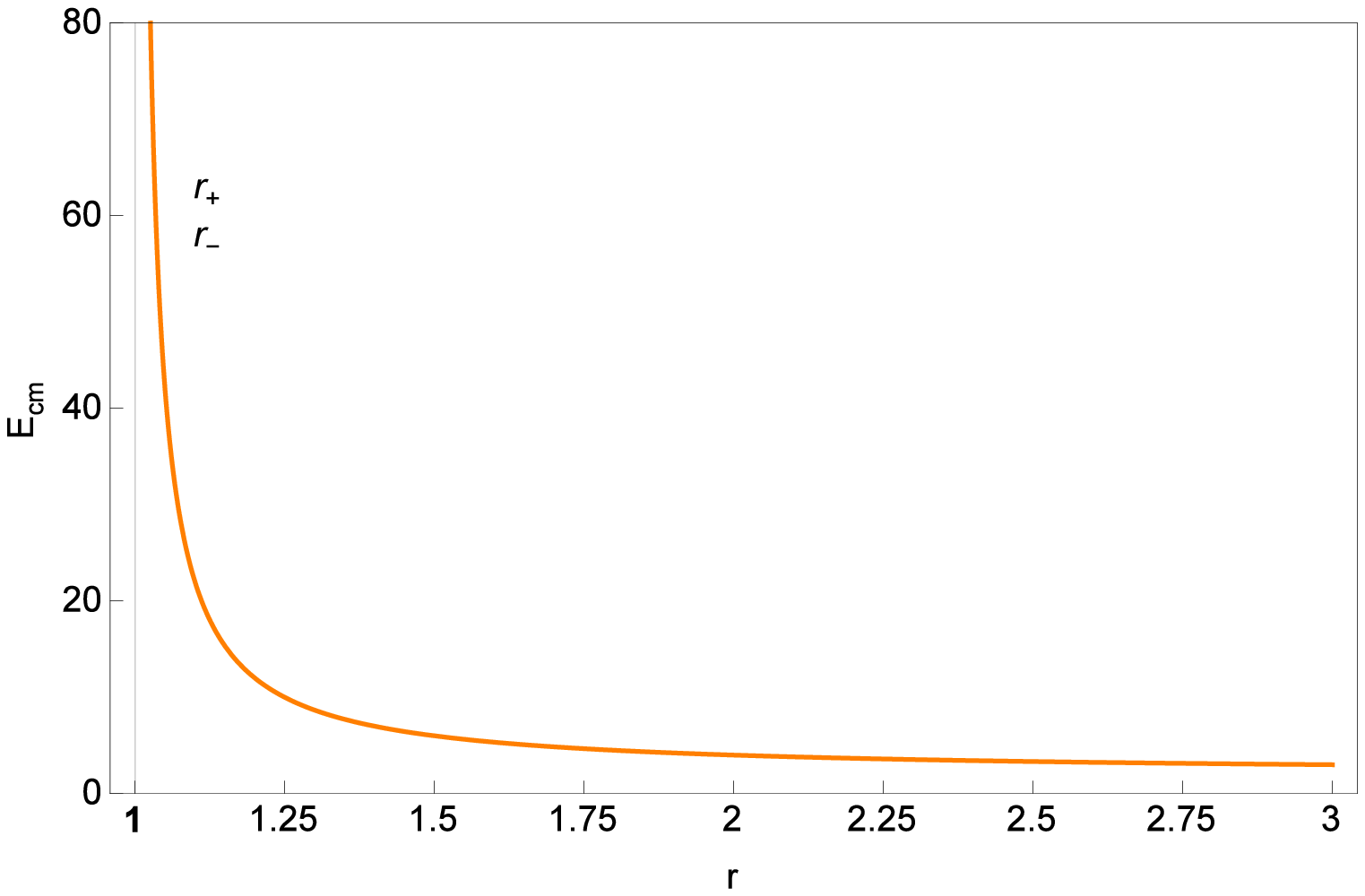}
\caption{The function $f(r)$ (left figure) and $E_{\text{cm}}$ (right figure) where first particle comes from far to the horizon of the extremal RN black hole with specific energy $E_{1}=1$ and second particle emanates from the white hole region with specific energy $E_{2}=1$. We set $\mu=1$, $m_{1}=m_{2}=1$ and $q=1$. Vertical line recognizes place of residence of the horizon.}\label{CMEf6}
\end{figure}
We plot the function $f(r)$ (left figures) for $\mu=1$. We set $q=0.7$, $q=0.75$, $q=0.8$, $q=0.85$, $q=0.9$, $q=1$ in Figures (\ref{CMEf1}), (\ref{CMEf2}), (\ref{CMEf3}), (\ref{CMEf4}), (\ref{CMEf5}) and (\ref{CMEf6}), respectively. We also plot $E_{\text{cm}}$ (right figures) where first particle comes from far to the outer horizon of the non-extremal RN black hole with mass $m_{1}=1$ and specific energy $E_{1}=1$ and second particle emanates from the white hole region with mass $m_{2}=1$ and specific energy $E_{2}=1$. Vertical lines recognize place of residence of the outer and inner horizons. Clearly, the $E_{\text{cm}}$ deviates at the different values of the outer horizon. Orange curves (\ref{CMEf6}) identify the function $f(r)$ and $E_{\text{cm}}$ for extreme case $r_{+}=r_{-}$.
\section{conclusion}
If two particles moving in a straight line directed opposite to each other in the spacetime then they can collide at any point. Collision of the particles occurs close to the outer horizon of the white hole. The second particle emanates from the white hole region should lapse near the bifurcation point but not crosses this point, because if it crosses this point, it would come into the $T-$region instead of $R-$region. There is no requirement of fine-tunning of parameters as compared to the BSW effect \cite{1}. There is only a kinematic restriction to attain unbounded $E_{\text{cm}}$ that is first particle moves in the $R-$region and second particle moves from a white hole to the $R-$region. Thus white holes and black holes can accelerate particles to extremely high energies.

\end{document}